\documentclass[aps,prl,floatfix,twocolumn,showpacs]{revtex4}
\usepackage{epsfig}
\usepackage{amsmath}
\begin{document}

\hsize\textwidth\columnwidth\hsize\csname@twocolumnfalse\endcsname

\title{Theory of phonon-induced spin relaxation in laterally coupled quantum dots}

\author{Peter Stano and Jaroslav Fabian}
\affiliation{Institute for Theoretical Physics, University of
Regensburg, 93040 Regensburg, Germany}

\vskip1.5truecm
\begin{abstract}
Phonon-induced spin relaxation in coupled lateral quantum dots in
the presence of spin-orbit coupling is calculated. The calculation
for single dots is consistent with experiment. Spin relaxation in
double dots at useful interdot couplings is dominated by spin hot
spots that are strongly anisotropic. Spin hot spots are ineffective
for a diagonal crystallographic orientation of the dots with a
transverse in-plane field. This geometry is proposed for spin-based
quantum information processing.

\end{abstract}
 \maketitle

Understanding spin relaxation in coupled quantum dots is important
for setting the efficiency of spin-based applications of information
processing, such as spin quantum computing \cite{Loss1998:PRA} or
controlled generation of spin entanglement \cite{Fabian2005:PRB}.
Phonon-induced spin relaxation has already been studied
theoretically in single dots for electrons \cite{Khaetskii2000:PRB, %
Khaetskii2001:PRB, Woods2002:PRB, Golovach2004:PRL, Cheng2004:PRB, %
Destefani2004:P, Bulaev2005:PRB, Falko2005:PRL, Sherman2005:PRB, %
Calero2005:PRL}, holes \cite{Bulaev2005:PRL}, and excitons
\cite{Tsitsishvili2005:PRB}, and in one-dimensional coupled dots
\cite{Romano2005:P}. Recently, spin relaxation of electrons in
single dots has been measured \cite{Elzerman2004:N}.

Here we present a realistic calculation of phonon induced spin
relaxation in single and coupled lateral quantum dots formed over a
depleted two dimensional electron gas in GaAs grown along [001], the
most typical growth direction. We  show that our calculation is
consistent with the single-dot experiment \cite{Elzerman2004:N}. We
predict that: (i) Spin relaxation in coupled dots is strongly
anisotropic with respect to the orientation of both an in-plane
magnetic field (due to the interplay of the Bychkov-Rashba and
Dresselhaus spin-orbit terms) and the dots' axis. The anisotropy is
limited by the in-plane inversion symmetry only. (ii) The spin
relaxation rate varies strongly with the inter-dot coupling, having
a giant enhancement at a significant range of useful tunneling
amplitudes due to spin hot spots (anticrossings caused by spin-orbit
coupling
\cite{Fabian1998:PRL, Fabian1999:PRL, Bulaev2005:PRB, %
Stano2005:PRB}). This variation, which is over several (four to
five) orders of magnitude, should be included in any realistic
modeling of spin coherence phenomena in coupled quantum dots with
controlled temporal evolution of the coupling. (iii) Fortunately,
the effects of (ii) are absent at specific configurations. The most
robust (with respect to materials parameters) such a configuration
is with dots oriented along [110] (or $[1\bar{1}0]$) with the
in-plane magnetic field along $[1\bar{1}0]$ ($[110]$). We propose to
use this configuration for spin-based quantum information
experiments.

Our single-electron Hamiltonian is $H = T + V + H_{SO} + H_Z$. Here
$T$ is the operator of the kinetic energy with the magnetic field
${\bf B}$ introduced by minimum coupling, and $V$ is the double-dot
confinement potential,
\begin{equation} \label{eq:conf}
V({\bf r}) = (1/2) m \omega_0^2 \min \{({\bf r} - {\bf d})^2, ({\bf
r} + {\bf d})^2 \}.
\end{equation}
The plane radius vector is ${\bf r}=(x,y)$, where $x=[100]$ and
$y=[010]$ are the crystallographic axes, while ${\bf d}$ defines the
distance (as well as tunneling energy at $B=0$) and the orientation
of the dots; the angle between ${\bf d}$ and $\hat{x}$ is denoted
below as $\delta$. The conduction electron mass is $m$ and the
single-dot ($d=0$) confining energy is $\hbar\omega_0$. The
spin-orbit coupling comprises three contributions
\cite{Zutic2004:RMP}: $H_{SO} = H_{BR} + H_{D} + H_{D3}$, where
\begin{eqnarray}
H_{BR} & = & \alpha_{BR} \left ( \sigma_x K_y - \sigma_y K_x \right ), \\
H_{D} & = & \gamma_c \langle \hat{K}_z^2 \rangle
\left  ( - \sigma_x K_x + \sigma_y K_y \right ), \\
H_{D3} & = & (\gamma_c/2) \left ( \sigma_x K_x K_y^2 - \sigma_y K_y
K_x^2 \right ) + {\rm h.c.},
\end{eqnarray}
are the Bychkov-Rashba, linear Dresselhaus, and cubic Dresselhaus
couplings. Kinematic wave vector operators are ${\bf K} = -i \nabla
+ (e/\hbar) {\bf A}$, where $\bf A$ is the vector potential to ${\bf
B}$. While both $\alpha_{BR}$ and the quantum average, $\langle
\hat{K}_z^2 \rangle$, in the growth direction $\hat{z}$, are tunable
by a top gate, $\gamma_c$ is a band parameter. Below we use $l_{BR}
= \hbar^2/2m\alpha_{BR}$ and $l_D = \hbar^2/2m\gamma_c \langle
\hat{K}_z^2 \rangle$ as effective spin-orbit lengths. The last term
in the Hamiltonian is the Zeeman splitting $ H_Z = - (g/2) \mu_B
\boldsymbol{\sigma} \cdot {\bf B}$, expressed by the band g-factor
$g$ and the Bohr magneton $\mu_B$.

\begin{figure}
\centerline{\psfig{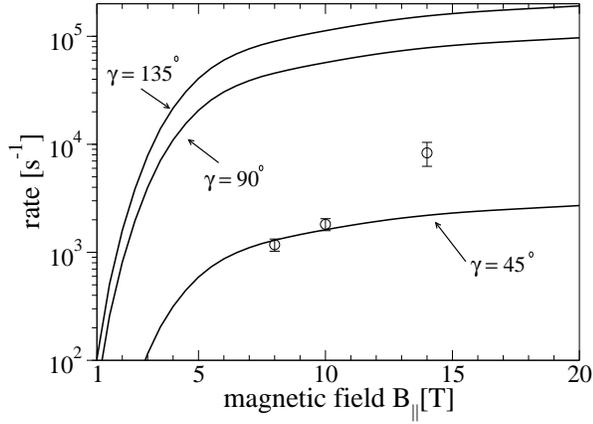}}
\caption{Calculated spin relaxation rate for a single quantum dot as
a function of $B_{||}$ applied along $[110]$, $[010]$, and
$[\bar{1}10]$. The symbols are experimental data from Ref.
\onlinecite{Elzerman2004:N}. The calculated curves for $B_{||} \agt
10 $ T are not realistic since they do not incorporate cyclotron
effects in the $z$ direction.} \label{fig:1}
\end{figure}

Single electron states are obtained by numerically diagonalizing
Hamiltonian $H$ using the Lanczos algorithm. The GaAs materials
parameters are used: $m= 0.067 m_e$ ($m_e$ is the free electron
mass), $g = -0.44$, and $\gamma_c = 27.5$ eV$\cdot
\AA^3$\cite{Zutic2004:RMP}. The linear Dresselhaus coupling is
chosen to be $\gamma_c \langle K_z^2 \rangle = 4.5$ meV$\cdot \AA$,
corresponding to a 11 nm wide ground state of a triangular confining
potential \cite{Sousa2003:PRB}. The Bychkov-Rashba parameter
$\alpha_{BR}$ is 3.3 meV$\cdot \AA$, in line with experiments
\cite{Miller2003:PRL, Knap1996:PRB}. The above  $\gamma_c \langle
K_z^2 \rangle$ and $\alpha_{BR}$ are selected to be both generic and
consistent with the experiment of Ref. \cite{Elzerman2004:N} (see
Fig. \ref{fig:1}). Our confining energy $\hbar \omega_0$ is $1.1$
meV, corresponding to the confining length of $l_0 = 32$ nm,
describing the experimental system of Ref. \cite{Elzerman2004:N}.
Finally, the magnetic vector potential is given in Landau's gauge,
${\bf A} = (B_\perp/2)(-y,x,0)$ for the case of a perpendicular
magnetic field ${\bf B} = B_\perp {\hat{ z}}$; if the field is in
plane, ${\bf B} = B_{||}(\cos\gamma, \sin\gamma,0)$, where $\gamma$
is the angle between the field and $\hat{x}$, cyclotron effects are
neglected. This is justified here for fields up to about 10 T, for
which the magnetic length is greater than the confining length in
the $z$-direction.

While spin-orbit terms couple opposite spin states, electron-phonon
coupling enables transitions between such states. Here we include
the most relevant terms---deformation and  piezoelectric acoustic
electron-phonon potentials (direct spin-dependent electron-phonon couplings
appear inefficient \cite{Khaetskii2000:PRB, Khaetskii2001:PRB}),
\begin{equation} \label{eq:ep}
H_{\rm ep}  =  \sum_{{\bf Q}\lambda} \sqrt{\frac{\hbar Q}{2\rho V
c_\lambda}}\Big(\sigma_e\delta_{l\lambda} -\frac{i eh_{14}
M_\lambda}{\sqrt{Q} }\Big)(b_{{\bf Q}\lambda} + b^{\dagger}_{-{\bf
Q}\lambda}) e^{i{\bf q}\cdot {\bf r}}.
\end{equation}
The summation is over phonon wave vectors ${\bf Q} = ({\bf q},Q_z)$
and polarizations $\lambda$ (two acoustic, $t$, and one
longitudinal, $l$). The phonon creation and annihilation operators
are denoted by $b^{\dagger}$ and $b$, respectively. For GaAs the
mass density is $\rho =5.3 \times 10^3$ kg/$m^{3}$, the phonon
velocities are $c_l = 5.3 \times 10^3$ m/s and $c_t = 2.5 \times
10^3$ m/s, the deformation potential $\sigma_e = 7.0$ eV, and the
piezoelectric constant $eh_{14} = 1.4 \times 10^9 $ eV/m; $V$ is the
unit cell volume and $N$ is the number of unit cells. The geometric
factors $M_\lambda$ depend only on the direction of ${\bf Q}$
\cite{Mahan:2000}.

\begin{figure}
\centerline{\psfig{file=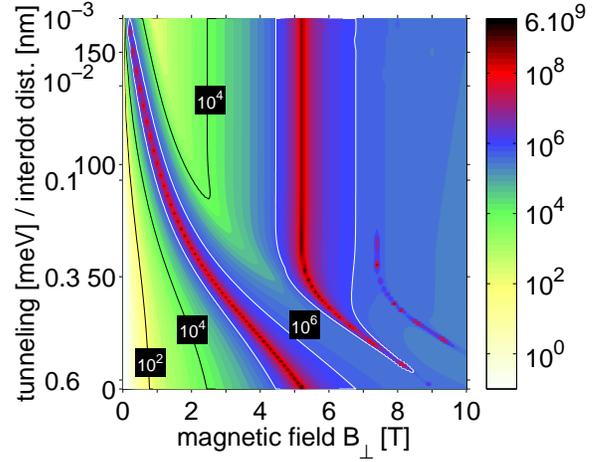,width=0.9\linewidth}}
\caption{(Color online) Calculated spin relaxation, in s$^{-1}$, for
a double quantum dot along [100] as a function of a perpendicular
magnetic field $B_\perp$ and tunneling energy/interdot distance. The
labeled contours are equirelaxational lines. Three spin hot spot
``ridges'' are visible with spin relaxation as large as $10^9$
s$^{-1}$. The granular pattern at spin hot spots is due to the
finite resolution of the graphics. } \label{fig:2}
\end{figure}

We calculate the rate of spin relaxation as the transition
probability (given by the Fermi golden rule) due to $H_{\rm ep}$,
from the upper Zeeman split ground state (denoted as
$\Gamma^{\downarrow}_S$ in \cite{Stano2005:PRB}) to {\it all} lower
states (which have necessarily opposite spin). If the Zeeman
splitting is smaller than the orbital excitation energy, the spin
relaxes to the ground state ($\Gamma_S^{\uparrow}$) only. If,
however, more orbital states are present below the upper Zeeman
split state, transitions to all lower states contribute to spin
relaxation. This is particularly relevant for spin relaxation in
single dots at large magnetic fields and in coupled dots at weak
couplings. The spin direction of a state is given by the sign of the
expectation value of $\boldsymbol{\sigma}$ in the direction of ${\bf
B}$.

In order to predict the spin relaxation rate in coupled dots, we
first discuss a single dot case and compare it with experiment. This
is shown in Fig. \ref{fig:1}, where spin relaxation as a function of
$B_{||}$ applied at different angles $\gamma$ is calculated. The
experiment specific value of $g= -0.35$ is taken. Spin-orbit
parameters could be adjusted from a fit to the experimental data.
However, such a fit is presently not possible since the calculated
rates depend strongly on $\gamma$, reflecting the reduced symmetry
($C_{2v}$) of the GaAs interface, while the experimental data are
taken for a single diagonal crystallographic direction, undetermined
whether [110] or $[1\overline{1}0]$ with respect to $C_{2v}$)
\cite{Vandersypen:PC}. These two directions are not equivalent,
which is reflected by the anisotropy shown in Fig. \ref{fig:1}. For
the purposes of demonstration we assume {\it ad hoc} that the
experiment is done for [110]; the spin-orbit parameters
used in this paper are selected (the selection is by no 
means unique) to quantitatively describe the
experiment with this $\gamma$ \cite{footnote1}. 
This is to demonstrate that the experiment is consistent
with the phonon-induced spin relaxation model for reasonable values
of spin-orbit parameters. We disagree with the experiment at 14 T in
which cyclotron effects (beyond the scope of our theory) in the
growth direction become important.

\begin{figure}
\centerline{\psfig{file=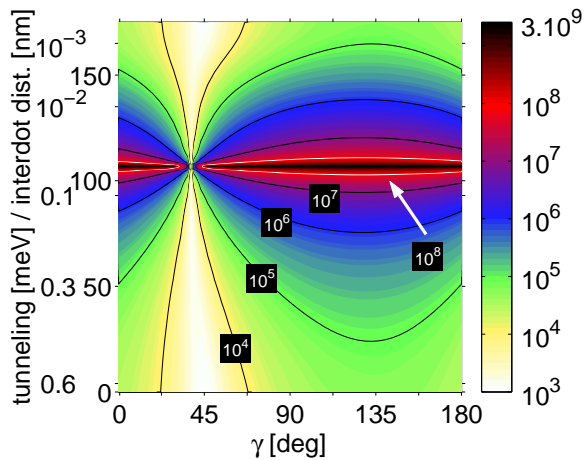,width=0.9\linewidth}}
\caption{(Color online) Calculated spin relaxation rate, in
s$^{-1}$, of a double quantum dot as a function of $\gamma$ and
tunneling energy, for $B_{||} = 5$ T. The dots are oriented along
[100]. The weakest relaxation is for $\gamma\approx 35^\circ$. Spin
hot spots strongly influence spin relaxation at tunneling energies
from 0.001 to 0.1 meV.} \label{fig:3}
\end{figure}

The calculated anisotropy in Fig. \ref{fig:1} appears due to the
reduced symmetry in the presence of both Bychkov-Rashba and
Dresselhaus terms \cite{Stano2005:PRB}. Using a unitary
transformation that eliminates the linear spin-orbit terms in a
confined system \cite{Aleiner2001:PRL, Stano2005:PRB} the Zeeman
term due to $B_{||}$ transforms to an effective Zeeman term with
magnetic field $B^{\rm eff}_{z}(x,y)\hat{z}$ along $z$:
\begin{eqnarray} \label{eq:Beff}
B^{\rm eff}_{z}=-B_{||}\left [ x \left (
 \frac{\cos \gamma}{l_{BR}} - \frac{\sin\gamma}{\l_D} \right ) + y \left
( \frac{\sin \gamma}{l_{BR}} - \frac{\cos\gamma}{l_D}\right) \right]
.
\end{eqnarray}
The spin-flip probability is proportional to the square of the
transition matrix element of $B^{\rm eff}_{z}$. Since the single dot
(Fock-Darwin) upper and lower Zeeman ground states are coupled
through the first excited orbital states which can be chosen to have
a definite $x$ or $y$ symmetry, the spin relaxation rate is
proportional to the sum of the squares at $x$ and $y$ in Eq.
\ref{eq:Beff}. The spin-flip probability is then proportional to the
inverse of the square of the effective anisotropic spin-orbit length
${\cal L}_{SO}(\gamma)$,
\begin{equation} \label{eq:Leff}
{\cal L}^{-2}_{SO}(\gamma) = 1/l_{BR}^2 + 1/\l_D^2 - 2\sin
(2\gamma)/l_{BR}l_D.
\end{equation}
The period of $\pi$ reflects the $C_{2v}$ symmetry of the interface
\cite{Stano2005:PRB}. The minimum spin relaxation is at $\gamma=
45^\circ$, while the maximum is at $135^\circ$, consistent with the
numerics in Fig. \ref{fig:1}. The anisotropy is absent if one of the
spin-orbit couplings dominates. Experimental observation of such an
anisotropy would be a clear signal of a phonon-induced spin
relaxation and could be used to extract the ratio of $l_D$ and
$l_{BR}$. If $l_{BR}=l_D$, the anisotropy is strongest---the spin
relaxation rate due to the linear spin-orbit terms vanishes for
$\gamma=45^\circ$. Details of the analytical derivations will be
published in a longer version of this article \cite{Stano:U}. The
anisotropy of Eq. \ref{eq:Leff} has been found earlier
\cite{Golovach2004:PRL}, while related anisotropies in g-factors has
been predicted for extended two-dimensional systems
\cite{Valin-Rodriguez2005:P}.

We now move to double quantum dots described by the confining
potential $V$ in Eq. \ref{eq:conf}. We have already predicted that
spin hot spots in these systems appear whenever the Zeeman splitting
equals the tunneling energy (difference between symmetric and
asymmetric orbital levels) \cite{Stano2005:PRB}. At weak coupling
($d\gg l_0$) the Zeeman splitting dominates and spin relaxation
proceeds through at least two channels, one to the symmetric
($\Gamma_S^{\uparrow}$), the other to the asymmetric
($\Gamma_A^{\uparrow}$) orbital state. At large coupling spin
relaxation is, in general, a single channel process, except at very
large magnetic fields in which Landau levels form. The two regimes
are separated by a spin hot spot in which spin relaxation is as
large as orbital relaxation. This regime is common---typical values
for the tunneling energy and the Zeeman splitting are of order 0.1
meV.

Consider first double dots with a perpendicular magnetic field
$B_\perp$ which contributes both the Zeeman splitting as well as
cyclotron effects. The calculated spin relaxation as a function of
tunneling energy and $B_\perp$ is shown in Fig. \ref{fig:2}. The
profile is rather complex. Spin relaxation is dominated by the
presence of spin hot spots which enhance spin relaxation in most
regimes of the control parameters. For tunneling energies below 0.2
meV the weakest spin relaxation is for magnetic fields from 2 to 5
T. At $d=0$ the calculated rate is that of single quantum dots.
There is a characteristic cusp structure as a function of $B_\perp$
at a spin hot spot at $B_\perp \approx 5$ T (such a cusp is not seen
for the in-plane field case in Fig.\ref{fig:1} since due to the
absence of cyclotron effects the spin-hot spots appear at large
fields of $B_{||} \approx 54$ T). Our calculation for this
single-dot case is in quantitative agreement with perturbative
calculations\cite{Bulaev2005:PRB}.

\begin{figure}
\centerline{\psfig{file=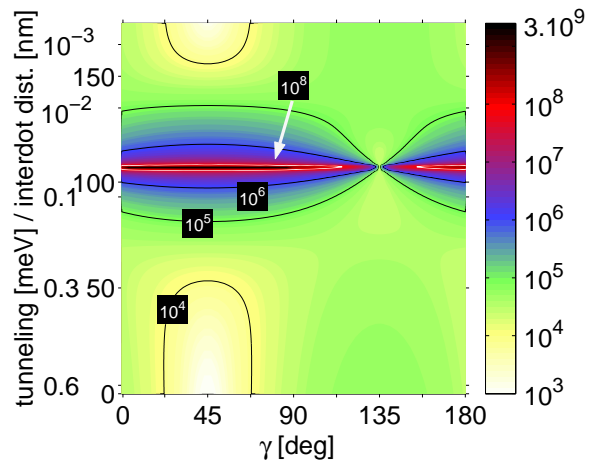,width=0.9\linewidth}}
\caption{(Color online) Calculated spin relaxation, in s$^{-1}$, of
a double dot as a function of $\gamma$ and the tunneling energy, for
$B_{||}=5$ T. The dots are oriented along [110]. The weakest spin
relaxation is for $\gamma = 135^\circ$, while at $\gamma=45^\circ$
spin hot spots do appear.} \label{fig:4}
\end{figure}

Let us now look at spin relaxation in an in-plane field $B_{||}$, a
situation interesting for spin qubit experiments, since cyclotron
effects are inhibited. Two important cases are shown for $B_{||}=5$
T and different orientations $\delta$. The first case, in which the
dots are along [100], is shown in Fig. \ref{fig:3}, and the second
case, in which the dots are along [110], is shown in Fig.
\ref{fig:4}. At small and large couplings, the spin relaxation rate
is strongly anisotropic, similar to the single-dot case in Fig.
\ref{fig:1}. This anisotropy is greatly enhanced in the intermediate
coupling by spin hot spots. In fact, spin hot spots dominate this
useful regime: spin relaxation is several orders of magnitude higher
than in the single dot case (limits of either very strong or very
weak coupling on the graph) at virtually all $\gamma$. The
exceptions are $\gamma \approx 35^\circ$ in Fig. \ref{fig:3} and
$\gamma = 135^\circ$ in Fig. \ref{fig:4}. At smaller (larger)
$B_{||}$, the strong relaxation regime moves towards smaller
(larger) coupling, while the two angles of ``easy passage'' remain.

The anisotropy in both $\gamma$ and $\delta$ can be explained by
transforming the effective Zeeman field $B^{\rm eff}_{z}$, Eq.
\ref{eq:Beff}, into the rotated coordinate system in which the $x$
axis lies along ${\bf d}$:
\begin{eqnarray} \label{eq:Beff1}
\tilde {B}^{\rm eff}_{z} & = & - B_{||} x\left [
l_{BR}^{-1}\cos(\gamma-\delta) - l_D^{-1}\sin(\gamma+\delta) \right
] \nonumber  \\
& +& B_{||} y \left [ l_{BR}^{-1}\sin(\gamma-\delta) -
l_D^{-1}\cos(\gamma+\delta) \right ].
\end{eqnarray}
Unlike in single dots, the relevant states in double dots are
coupled by $x$ and $y$ differently. At weak coupling the dominant
term is the one containing $x$ (which is the symmetry of the first
excited orbital state $\Gamma_A$ \cite{Stano2005:PRB}). This term leads to anisotropic spin hot
spots and giant spin relaxation seen in Figs. \ref{fig:3} and \ref{fig:4}. Spin hot
spots vanish if $l_{BR}^{-1}\cos(\gamma-\delta) -
l_D^{-1}\sin(\gamma+\delta)=0$, which, for $\delta=0$, gives $\tan
\gamma = l_D/l_{BR}$. For our parameters the corresponding angle is
about 35$^\circ$, consistent with the numerical calculation shown in
Fig. \ref{fig:3}. Unfortunately, this angle depends on the
spin-orbit coupling and is thus not robust against materials and
growth details. On the other hand, for $\delta = 45^\circ$, the spin
hot spots vanish if $\gamma = 135^\circ$, which is a universal value
independent of spin-orbit coupling. This is confirmed numerically in
Fig. \ref{fig:4}. In the special case of $l_{BR} = l_D$, the
condition for the weakest spin relaxation would be
$\gamma=45^\circ$, as in single dots.

In GaAs single dots spin hot spots, which appear at large magnetic
fields, are due to the Bychkov-Rashba coupling only
\cite{Bulaev2005:PRB,Stano2005:PRB}. On the other hand, as can be
read from Eq. \ref{eq:Beff1},  in double dots spin hot spots appear
at {\it arbitrary} small magnetic fields and are caused by {\it
both} the Bychkov-Rashba and Dresselhaus couplings
\cite{Stano2005:PRB}, whose interference causes spin hot spot
anisotropy.

Similar results apply for other growth directions. In [111] $H_{D}
\sim H_{BR}$ \cite{Zutic2004:RMP} and the above results in the limit
$\l_D \to \infty$ apply (with $l_{BR}$ being a combination of both
coupling strengths). The spin hot spots are inhibited for any
orientation of double dots and a perpendicular in-plane field
[$\cos(\gamma - \delta) =0$]; such a configuration can also be used
in applications. In the [110] case a unique easy
passage exists for $\gamma=0$ and $\delta = 90^\circ$
\cite{Stano:U}.

In conclusion, we have performed realistic calculations of
phonon-induced spin relaxation in double quantum dots in the
presence of magnetic field. The spin relaxation rate is dominated by
spin hot spots in the useful regime of interdot couplings. The spin
hot spot anisotropy allows an inhibited spin relaxation for the dots
oriented along a diagonal of the [001] plane with a
transverse in-plane magnetic field.

We thank U. R\"ossler for useful discussions. This work was
supported by the US ONR.

\bibliography{references,footnotes}

\begin{thebibliography}{29}
\bibitem[{foo()}]{footnote1}
\bibinfo{note}{If we took the $[1\overline{1}0]$ direction to correspond to the
  experimental situation, the spin-orbit parameters would need to have values
  at the lower limit of what appears to be realistic for GaAs systems.}
\end{thebibliography}

\end{document}